\documentclass{elsart}
\usepackage{graphicx,amssymb,natbib}
\begin{document}

\newcommand{\smsub}[2]{{#1}_{\rm #2}}
\newcommand{\mqg}{\smsub{m}{QG}}
\newcommand{\mgut}{\smsub{m}{GUT}}
\newcommand{\mpl}{\smsub{m}{Pl}}
\newcommand{\mnuc}{\smsub{m}{nuc}}
\newcommand{\mh}{\smsub{m}{H}}
\newcommand{\bqcd}{\smsub{\beta}{QCD}}

\begin{frontmatter}
\title{The Science Case for STEP}
\author{James Overduin,$^1$ Francis Everitt,$^2$ John Mester$^3$ and
   Paul Worden$^4$}
\address{Gravity Probe~B, Hansen Experimental Physics Laboratory,\\
   Stanford University, Stanford, CA 94305, USA}
\thanks[footnote1]{{\tt overduin@relgyro.stanford.edu}}
\thanks[footnote2]{{\tt francis@relgyro.stanford.edu}}
\thanks[footnote3]{{\tt mester@relgyro.stanford.edu}}
\thanks[footnote4]{{\tt worden@relgyro.stanford.edu}}
\begin{abstract}
STEP (the Satellite Test of the Equivalence Principle) will advance
experimental limits on violations of Einstein's equivalence principle (EP)
from their present sensitivity of 2~parts in $10^{13}$ to 1~part in $10^{18}$
through multiple comparison of the motions of four pairs of test masses of
different compositions in an earth-orbiting drag-free satellite.
Dimensional arguments suggest that violations, if they exist, should be found
in this range, and they are also suggested by leading attempts at unified
theories of fundamental interactions (e.g. string theory) and cosmological
theories involving dynamical dark energy. Discovery of a violation would
constitute the discovery of a new force of nature and provide a critical
signpost toward unification. A null result would be just as profound, because
it would close off any possibility of a natural-strength coupling between
standard-model fields and the new light degrees of freedom that such theories
generically predict (e.g., dilatons, moduli, quintessence).  STEP should thus
be seen as the intermediate-scale component of an integrated strategy for
fundamental physics experiments that already includes particle acclerators
(at the smallest scales) and supernova probes (at the largest).
The former may find indirect evidence for new fields via their missing-energy
signatures, and the latter may produce direct evidence through changes in
cosmological equation of state---but only a gravitational experiment like
STEP can go further and reveal how or whether such a field couples to the rest
of the standard model. It is at once complementary to the other two kinds of
tests, and a uniquely powerful probe of fundamental physics in its own right.
\end{abstract}
\end{frontmatter}

\section{Historical Overview}

The Satellite Test of the Equivalence Principle (STEP; Fig.~1) 
\begin{figure}
\begin{center}
\includegraphics*[width=10.7cm]{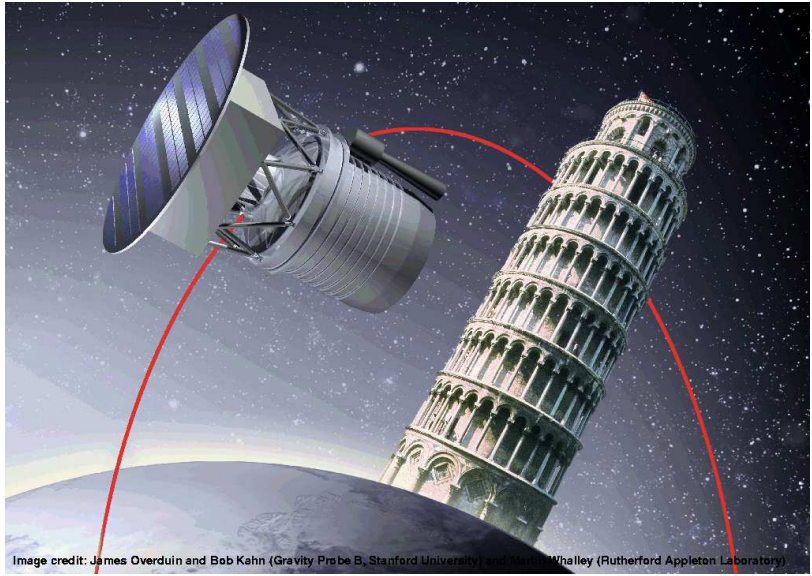}
\end{center}
\caption{Artists' impression of the STEP spacecraft}
\end{figure}
will probe the foundation of Einstein's theory of general relativity,
the (local) equivalence of gravitational and inertial mass---more
specifically called the weak equivalence principle---to unprecedented
precision.  The equivalence principle (EP) originated in Newton's clear
recognition (1687) of the strange experimental fact that mass fulfills
two conceptually independent functions in physics, as both the source of
gravitation and the seat of inertia.  Einstein's ``happiest thought'' (1907)
was the recognition that the equivalence of gravitational and inertial mass
allows one to locally ``transform away'' gravity by moving to the same
accelerated frame, regardless of the mass or composition of the falling object.
It followed that the phenomenon of gravitation could not depend on any
property of matter, but must rather spring from some property of spacetime
itself.  Einstein identified the property of spacetime that is responsible
for gravitation as its curvature.  General relativity, our currently accepted
``geometrical'' theory of gravity, thus rests on the validity of the EP.
But it is now widely expected that general relativity must break down at
some level in order to be united with the other fields making up the
standard model of particle physics.  It therefore becomes imperative to test
the EP as carefully as possible.

Historically, equivalence has been tested in four distinct ways:
(1)~Galileo's free-fall method, (2)~Newton's pendulum experiments, (3)~Newton's
celestial method (his dazzling insight that moons and planets could be used
as test masses in the field of the sun) and (4)~E\"otv\"os' torsion balance.
Certain kinds of EP violation can also be constrained by other phenomena
such as the polarization of the cosmic microwave background \citep{ni08}.
However, the most robust and sensitive EP tests to date have come from
approaches (3) and (4).  The celestial method now makes use of lunar laser
ranging to place limits on the relative difference in acceleration toward the
sun of the earth and moon of $(-1.0\pm1.4)\times 10^{-13}$ \citep{williams04},
and modern state-of-the-art torsion balance experiments give comparable
constraints of $(0.3\pm1.8)\times 10^{-13}$ \citep{schlamminger08}.  Both
these methods have reached an advanced level of maturity and it is unlikely
that they will advance significantly beyond the $10^{-13}$ level in the near
term.

STEP is conceptually a return to Galileo's free-fall method, but one that
uses a 7000~km high ``tower'' that constantly reverses its direction to give
a continuous periodic signal, rather than a quadratic 3~s drop
(Fig.~2). 
\begin{figure}
\begin{center}
\includegraphics*[width=8.4cm]{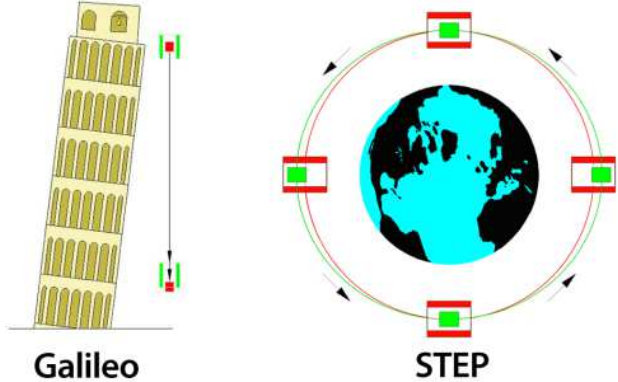}
\end{center}
\caption{Compared with Galileo at Pisa, STEP employs a ``tower'' not 60~m but
   7000~km high, and one constantly reversing its direction to give a 
   continuous periodic signal rather than a quadratic 3~s drop.}
\end{figure}
A free-fall experiment in space has two principal advantages over terrestrial
torsion-balance tests: a larger driving acceleration (sourced by the entire
mass of the earth) and a quieter ``seismic'' environment, particularly if
drag-free technology is used.  These factors alone lead to an increase in
sensitivity over current methods of three orders of magnitude each.
Comprehensive numerical disturbance analysis \citep{worden01} has
established that 20 orbits (1.33 days) of integration time will suffice
for STEP to improve on existing EP constraints by {\em five to six orders
of magnitude\/}, from $\sim\!10^{-13}$ to $10^{-18}$.

There are other proposed
experiments to test the EP in space; most notable is MicroSCOPE, a
room-temperature mission that will fly two accelerometers with a measurement
goal of $10^{-15}$.  MicroSCOPE was conceived by the ONERA group (part of the
STEP collaboration) and is funded by CNES and ESA with a possible launch date
of 2011.  This paper will focus on STEP, which while not yet approved for
flight, has been through Phase~A studies sponsored by NASA and ESA, and
promises a large improvement in sensitivity.

\section{Experimental Design}

The STEP design calls for four pairs of concentric test masses,
currently composed of Pt-Ir alloy, Nb and Be in a cyclic condition
to eliminate possible sources of systematic error
(the total acceleration difference between A-B, B-C and C-A must be zero
for three mass pairs AB, BC and CA).  This choice of test-mass materials
is not yet fixed, but results from extensive theoretical discussions
suggesting that EP violations are likely to be tied to three potential
determinative factors that can be connected to a general class of 
string-inspired models: baryon number, neutron excess and nuclear
electrostatic energy [Fig.~3; 
\cite{damour94,damour96,blaser01}].
\begin{figure}[t!]
\begin{center}
\includegraphics*[width=10.4cm]{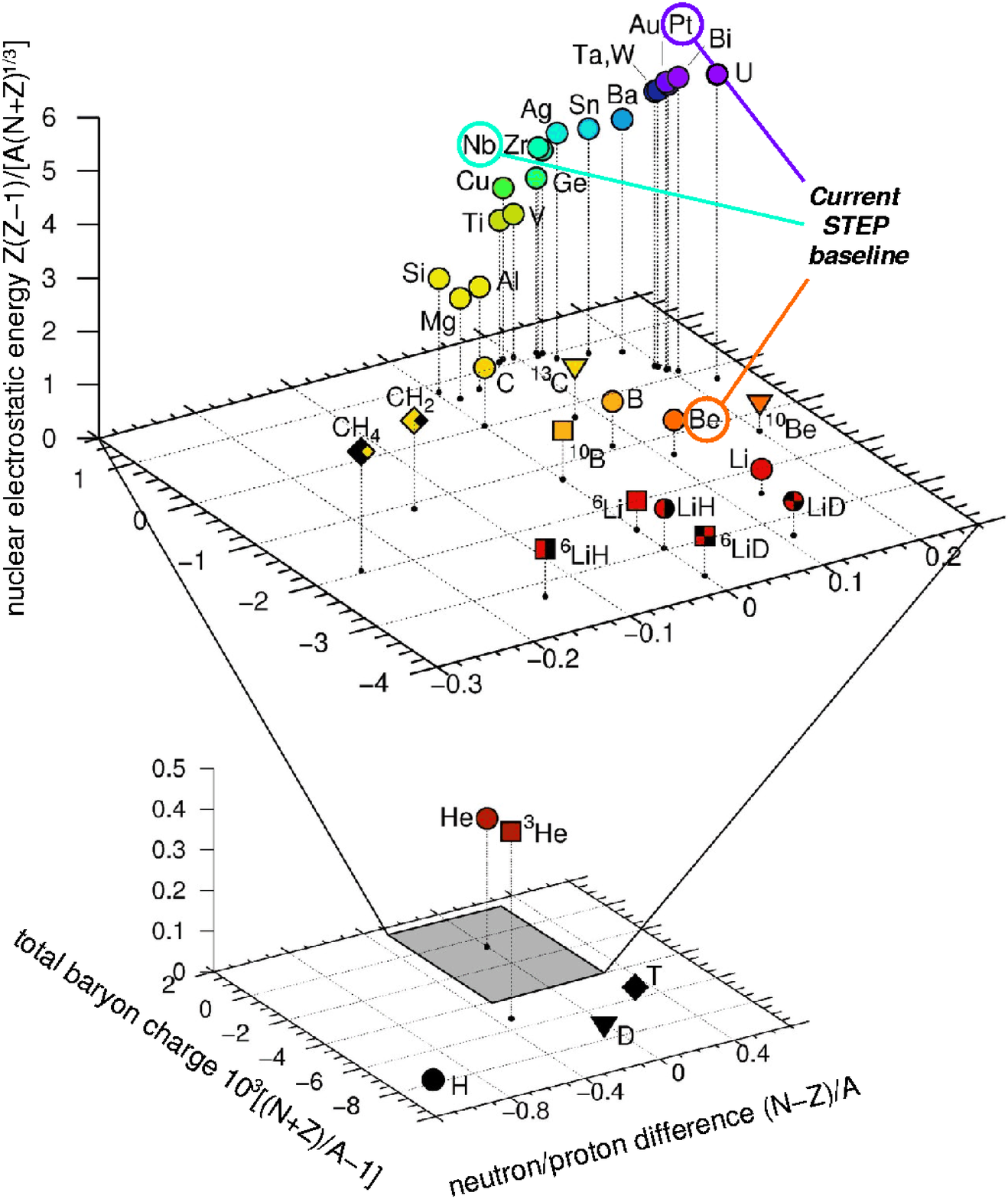}
\end{center}
\caption{Test mass choice.  A theoretical approach might be to span
the largest possible volume in the space of string-inspired ``elementary
charges'' such as baryon number $N+Z$, neutron excess $N-Z$ and nuclear
electrostatic energy $[\propto Z(Z-1)]$, all normalized by atomic mass $A$
(see text).  This approach must be balanced against practical issues such
as manufacturability, cost, and above all the need to make sure that
anything measured is a real effect.}
\end{figure}
The test masses are constrained by superconducting magnetic bearings to move
in one direction only; they can be perfectly centered by means of gravity
gradient signals, thus avoiding the pitfall of most other free-fall methods
(unequal initial velocities and times of release).  Their accelerations are
monitored with very soft magnetic ``springs'' coupled to a cryogenic
SQUID-based readout system.  The SQUIDs are inherited from Gravity Probe~B,
as are many of the other key STEP technologies, including test-mass caging
mechanisms, charge measurement and UV discharge systems, drag-free control
algorithms and proportional helium thrusters using boiloff from the dewar
as propellant (Fig.~4). 
\begin{figure}[t!]
\begin{center}
\includegraphics*[width=11.5cm]{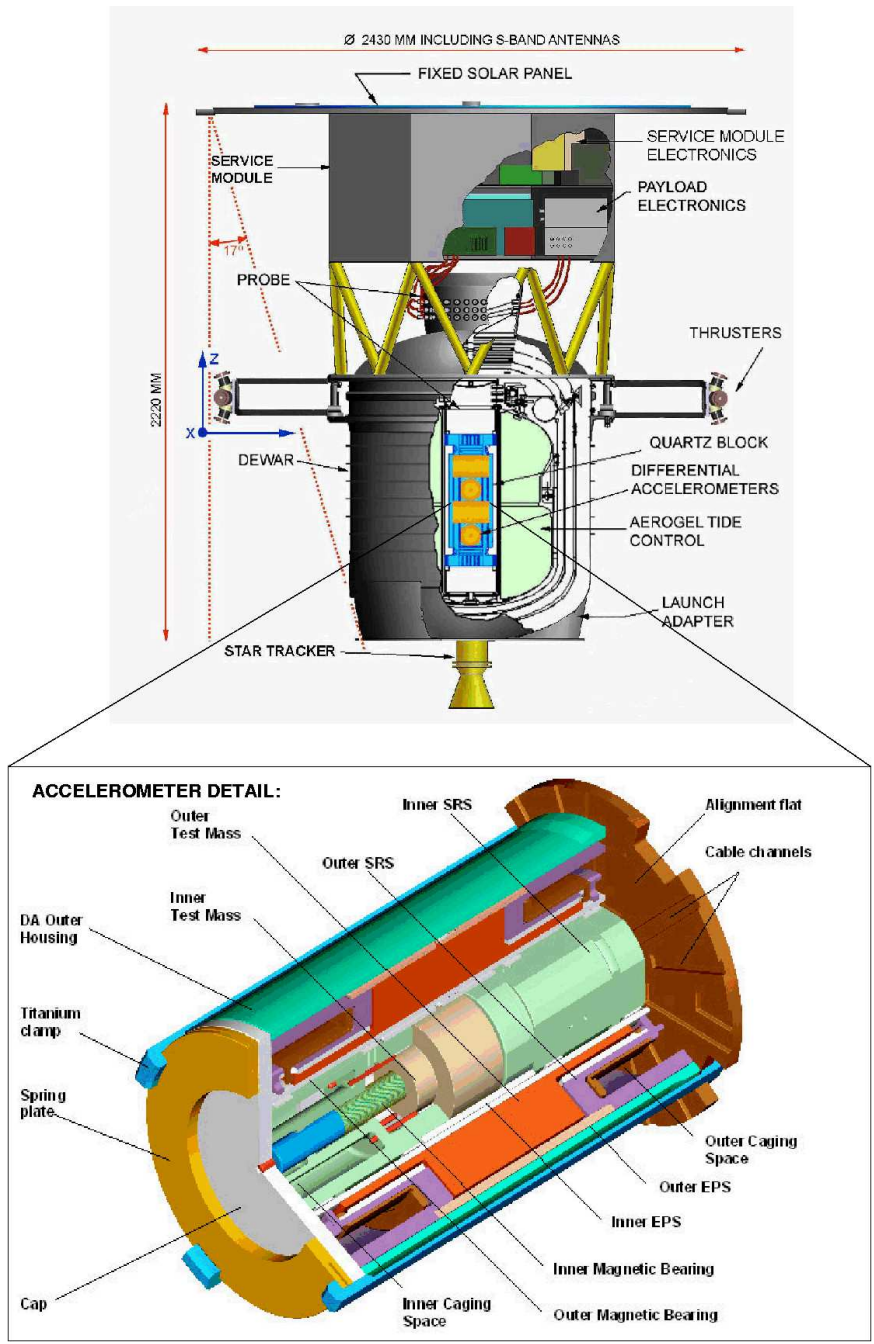}
\end{center}
\caption{Cutaway view of the STEP spacecraft (top) with accelerometer detail
   (bottom; DA=Differential Accelerometer, EPS=Electrostatic Positioning
   System, SRS=Squid Readout Sensor).  Each accelerometer contains two
   concentric test masses, cylindrical in shape but with dimensions
   chosen to make them gravitationally ``spherical'' to sixth order
   in mass moments.}
\end{figure}
Prototypes of key components including the accelerometer are in advanced
stages of development.

\section{Theoretical Motivation}

Theoretically, the range $10^{-18}\lesssim\Delta a/a\lesssim 10^{-13}$ is an
extremely interesting one.  This can be seen in at least three ways.  The
simplest argument is a dimensional one.  New effects in any theory of quantum
gravity must be describable at low energies by an effective field theory with
new terms like $\beta(m/\mqg)+O(m/\mqg)^2$ where $\beta$ is a 
dimensionless coupling parameter not too far from unity and $\mqg$ is the
quantum-gravity energy scale, which could be anywhere between the grand
unified theory (GUT) scale $\mgut\sim10^{16}$~GeV and the Planck scale
$\mpl\sim10^{19}$~GeV.  In a theory combining gravity with the standard
model of particle physics, $m$ could plausibly lie anywhere between the
mass of an ordinary nucleon ($\mnuc\sim1$~GeV) and that of the Higgs boson
($\mh\sim100$~GeV).  With these numbers one finds that EP-violating effects
should appear between $(\mnuc/\mpl)\sim 10^{-19}$ and
$(\mh/\mgut)\sim 10^{-14}$---exactly the range of interest.
\cite{adler06} has noted that this makes STEP a potential probe of
quantum gravity.

The dimensional argument, of course, is not decisive.  A second approach
is then to look at the broad range of specific theories that are
sufficiently mature to make quantitative predictions for EP violation.
There are two main categories.  On the high-energy physics side, EP violations
occur in many of the leading unified theories of fundamental interactions,
notably {\em string theories\/} based on extra spatial dimensions.
In the low-energy limit, these give back classical general relativity with
a key difference: they generically predict the existence of a
four-dimensional scalar dilaton partner to Einstein's tensor graviton,
and several other gravitational-strength scalar fields known as moduli.
In the early universe, these fields are naturally of the same order as the
gravitational field, and some method has to be found to get rid of them in
the universe we observe.  If they survive, they will couple to standard-model
fields with the same strength as gravity, producing drastic violations of the
EP.  One conjecture is that they acquire large masses and thus correspond to
very short-range interactions, but this solution, though widely accepted,
entails grave difficulties (the Polonyi or ``moduli problem'') because the
scalars are so copiously produced in the early universe that their masses
should long ago have overclosed the universe, causing it to collapse.
Another possibility involves a mechanism whereby a massless ``runaway dilaton''
(or moduli) field is cosmologically attracted toward values where it almost,
but not quite, decouples from matter; this results in EP violations that
overlap the range identified above and might in principle reach the
$\sim\!10^{-12}$ level \citep{damour02}.  Similar comments apply to another
influential model, the TeV ``little string'' theory \citep{antoniadis01}.

The second category of specific EP-violating theories occurs at the opposite
extremes of mass and length, in the field of cosmology.  The reason, however,
is the same: a new field is introduced whose properties are such that
it should naturally couple with gravitational strength to standard-model
fields, thus influencing their motion in violation of the EP.  The culprit in
this case is typically {\em dark energy\/}, a catch-all name for the surprising
but observationally unavoidable fact that the expansion of the universe appears 
to be undergoing late-time acceleration.  Three main explanations have been
advanced for this phenomenon: either there is a cosmological constant (whose
value is extremely difficult to understand), or general relativity is incorrect
on the largest scales---or dark energy is {\em dynamical\/}.  Most theories
of dynamical dark energy (also known as quintessence) involve one or more
species of new, light scalar fields that could violate the EP \citep{bean05}.
The same thing is true of new fields that may be responsible for producing
cosmological variations in the values of fundamental constants such as the
electromagnetic fine-structure constant $\alpha$ \citep{dvali02}.

In all or most of these specific theories, EP violations typically appear
within the STEP range, $10^{-18}\lesssim \Delta a/a\lesssim 10^{-13}$.
To understand the reasons for this, it is helpful to look at the third of
the arguments alluded to above for regarding this range as a particularly
rich and interesting one from a theoretical point of view.  This line of
reasoning shares some of the robustness of the dimensional argument, in
that it makes the fewest possible assumptions beyond the standard model,
while at the same time being based upon a convincing body of detailed
calculations.  Many authors have done work along these lines, with perhaps
the best known being that of \cite{carroll98} and \cite{chen05}, which we
follow in outline here.  Consider the simplest possible new field: a scalar
$\phi$ (as motivated by observations of dark energy, or alternatively by the
dilaton or supersymmetric moduli fields of high-energy unified theories such
as string theory).  Absent some protective symmetry (whose existence would
itself require explanation), this new field $\phi$ couples to standard-model
fields via dimensionless coupling constants $\beta_k$ (one for each
standard-model field) with values not too far from unity.  Detailed
calculations within the standard model (modified only to incorporate $\phi$)
show that these couplings are tightly constrained by existing limits on
violations of the EP.  The current bound of order $\Delta a/a<10^{-12}$
translates directly into a requirement that the dominant coupling factor
(the one associated with the gauge field of quantum chromodynamics or QCD)
cannot be larger than $\bqcd<10^{-6}$.  This is very small for a dimensionless
coupling constant, though one can plausibly ``manufacture'' dimensionless
quantities of this size (e.g. $\alpha^2/16\pi$), and many theorists would
judge that anything smaller
is almost certainly zero.  Now STEP will be sensitive to violations as small
as $10^{-18}$.  If none are detected at {\em this\/} level, then the
corresponding upper bounds on $\bqcd$ go down like the square root of
$\Delta a/a$; i.e., to $\bqcd<10^{-9}$, which is no longer a natural coupling
constant by any current stretch of the imagination.  For perspective, recall
the analogous ``strong CP'' problem in QCD, where a dimensionless quantity
of order $10^{-8}$ is deemed so unnatural that a new particle, the axion,
must be invoked to drive it toward zero.  This argument does not say that
EP violations inside the STEP range are inevitable; rather it suggests that
violations {\em outside\/} that range would be so unnaturally fine-tuned as
to not be worth looking for.  As Ed Witten has stated, ``It would be
surprising if $\phi$ exists and would not be detected in an experiment that
improves bounds on EP violations by 6 orders of magnitude'' \citep{witten00}.
{\em Only a space test of the EP has the power to force us to this conclusion.}

The fundamental nature of the EP makes such a test a win-win proposition,
regardless of whether violations are actually detected.  A positive detection
would be equivalent to the discovery of a new force of nature, and our first
signpost toward unification.  A null result would imply either that no
such field exists, or that there is some deep new symmetry that prevents its
being coupled to SM fields.  A historical parallel to a null result might be
the Michelson-Morley experiment, which reshaped physics because it found
nothing.  The ``nothing'' finally forced physicists to accept the fundamentally
different nature of light, at the cost of a radical revision of their concepts
of space and time.  A non-detection of EP violations at the $10^{-18}$ level
would strongly suggest that gravity is so fundamentally different from the
other forces that a similarly radical rethinking will be necessary to
accommodate it within the same theoretical framework as the SM based on
quantum field theory.

STEP should be seen as the integral intermediate-scale element of a
concerted strategy for fundamental physics experiments that already includes
high-energy particle accelerators (at the smallest scales) and cosmological
probes (at the largest scales), as suggested in Fig.~5. 
\begin{figure}[t!]
\begin{center}
\includegraphics*[width=13.9cm]{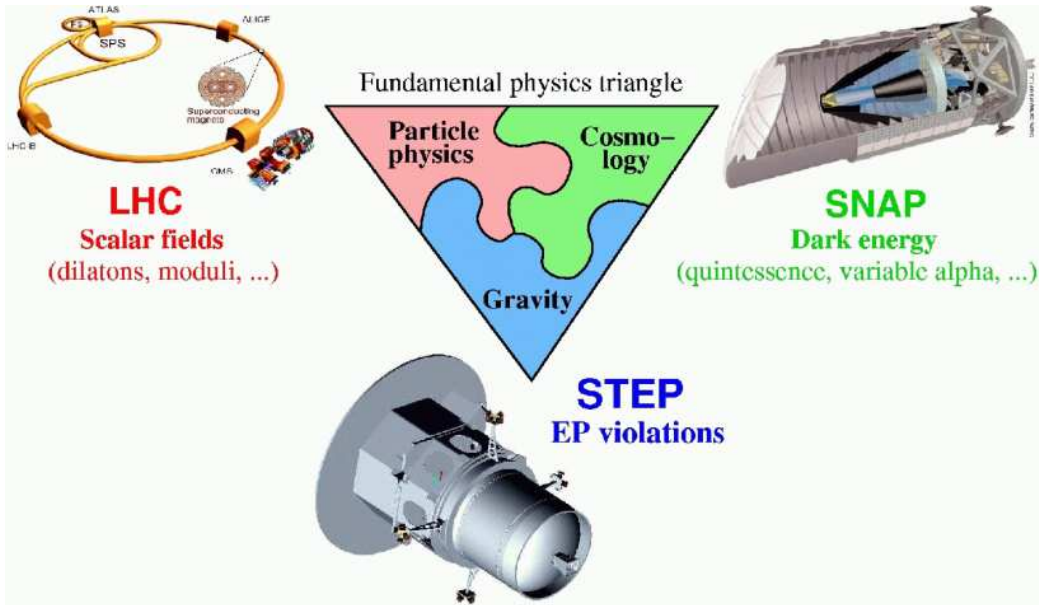}
\end{center}
\caption{Investigating nature on all three scales: small, large---{\em and
   intermediate}}
\end{figure}
Accelerators such as the Large Hadron Collider (LHC)
may provide indirect evidence for the existence of new fields
via their missing-energy signatures.  Astronomical observatories such as the
SuperNova Acceleration Probe (SNAP) may produce {\em direct\/} evidence of
a quintessence-type cosmological field through its bulk equation of state.
But {\em only a gravitational experiment such as STEP can go further and
reveal how or whether that field couples to the rest of the standard model.}
It is at once complementary to the other two kinds of tests, and a uniquely
powerful probe of fundamental physics in its own right.

\section*{Acknowledgments}

Thanks go to Sean Carroll for enlightening discussions.

\end{document}